# Experimental quantum secure direct communication with single photons

Jianyong Hu[1], Bo Yu[1], Mingyong Jing[1], Liantuan Xiao[1], Suotang Jia[1], Guoqing Qin[2,3,4] and Guilu Long[2,3,4]

[1]State Key Laboratory of Quantum Optics and Quantum Optics Devices, Institute of Laser Spectroscopy, Shanxi University, Taiyuan 030006, ShanXi province, China; [2]State Key Laboratory of Low-Dimensional Quantum Physics and Department of Physics, Tsinghua University, Beijing 100084, China; [3]Collaborative Innovation Center of Quantum Matter, Beijing 100084, China; [4]Tsinghua National Laboratory for Information Science and Technology, Tsinghua University, Beijing 100084, China.

Abstract: Quantum communication holds promise for absolutely security in secret message transmission. Quantum secure direct communication is an important mode of the quantum communication in which secret messages are securely communicated over a quantum channel directly. It has become one of the hot research areas in the last decade, and offers both high security and instantaneousness in communication. It is also a basic cryptographic primitive for constructing other quantum communication tasks such as quantum authentication, quantum dialogue and so on. Here we report the first experimental demonstration of quantum secure direct communication with single photons. The experiment is based on the DL04 protocol, equipped with a simple frequency coding. It has the advantage of being robust against channel noise and loss. The experiment demonstrated explicitly the block data transmission technique, which is essential for quantum secure direct communication. In the experiment, a block transmission of 80 single photons was demonstrated over fiber, and it provides effectively 16 different values, which is equivalent to 4 bits of direct transmission in one block. The experiment has firmly demonstrated the feasibility of quantum secure direct communication in the presence of noise and loss.

## INTRODUCTION

Secure communication is not only vital in military use and national security, but also important in modern everyday life. Quantum communication provides a novel way of communication with provable security. There are different modes of quantum communication, where each mode completes a specific task. Quantum key distribution (QKD) is the earliest and widely studied mode,[1] where two remote users distribute securely a random key. In quantum secret sharing,[2] a secret key is shared among two or more agents through quantum channels, and the key could only be constructed if the entire agents collaborate. With quantum secure direct communication (QSDC),[3] secret messages can be communicated directly through a quantum channel without first establishing a key. In quantum teleportation,[4] an unknown quantum state can be sent to a distant agent without actually sending the particle physically. In quantum dense coding,[5] two bits of information can be transmitted by just sending a particle with only a single bit. A fundamental difference between quantum and classical communication is the capability to detect eavesdropping on-site. Classical communication has no means to find eavesdropping during its operation. In contrast, quantum communication can detect Eve on-site, because according to quantum principle, any eavesdropping will cause change in the states and leads a sharp increase in the error rate. During a QKD process, sampling measurements are made constantly to estimate the error rate. If the error rate is lower than a given threshold, then the communication is concluded safe and the transmitted data are retained as raw key. If the error rate is higher than the threshold, the communication is considered insecure and the transmitted data are discarded. Because this information leakage before detection of Eve, it is essential that data transmitted in QKD be random, otherwise information will be leaked. Experimental development of QKD has been fast, and has reached a distance of a few hundred kilometers.[6]

In some QKD protocols, such as the BB84 or BBM92 protocols,[1,7] the random key is distributed indeterminately. In the determinate QKD protocols,[8-12] one user, Alice can first generate a random key, and then deterministically sends it to the other user, Bob. Though there is a difference in distributing the key, their key distribution nature does not change. The data transmitted must be random, otherwise the data transmitted before Eve's detection will be divulged. After the key is established, it is used to encrypt the message into ciphertext and then be transmitted through a classical communication to the other party. With deterministic QKD, the secret message transmission can be varied. Namely, Alice first chooses a random key and uses it to encrypt the secret message into ciphertext. Then Alice transmits the ciphertext to Bob through a quantum channel using the deterministic QKD. If they are certain that no eavesdropper exists, Alice sends the key through a classical channel to Bob. This is often called deterministic QKD communication, which is essentially a determinate QKD process plus a classical communication.[13]

In contrast to QKD, QSDC sends secret information directly through a quantum channel with provable security without setting up a key first,[3] which is a great improvement to the classical communication mode.[14] Like QKD, the security of QSDC also relies on quantum principles such as the no-cloning theorem, the uncertainty principle, correlation of entangled particles and nonlocality and so on. In addition, QSDC uses a block transmission technique, proposed in Ref. 3. In the block transmission,[3] the quantum information carriers such as EPR pairs or single photons are transmitted in block of $N$ particles. Then eavesdropping check is performed on the block so as to determine if there exists eavesdropping. Block transmission not only detects eavesdropping, but also avoids the leakage of information before detection of Eve,[3] hence enabling the direct secure communication. Of course, QSDC can also transmit random numbers and be used as QKD. After the first QSDC protocol (the efficient-QSDC protocol[3]) was proposed; many other QSDC protocols have been constructed. In 2003, Deng, Long and Liu proposed the two-step QSDC protocol[15] where the criteria for QSDC were explicitly stated. QSDC protocols with high-dimensional entanglement,[16-18] with multipartite entanglement,[19-21] with hyperentanglement[22] were developed. However, entanglement is not necessary for quantum communication. The first QSDC protocol with single photons was proposed in Ref. 23, the so-called DL04 protocol. Afterwards, many QSDC protocols based on single photons were proposed.[24-26] The secure direct nature of QSDC makes it an important cryptographic primitive. Protocols of quantum signature,[27] quantum dialogues,[28,29] and quantum direct secret sharing[30,31] were all constructed using QSDC. QSDC has become one of the hot research topics in quantum communication in the last decade.

The DL04 protocol is the first QSDC protocol based on single photon,[23] which is easier to implement than those with entanglement sources. It has been experimentally demonstrated in some special cases, namely the $N$=1 case of the DL04 protocol, denoted as DL04N1, or LM05, which is a robust four-state two-way deterministic QKD protocol.[32,33,34] Based on DL04N1, Deng et al constructed a circular quantum secret sharing protocol,[35] which has been experimentally demonstrated by Jin et al over 50 km in fibers.[36] DL04N1 protocol has been proved to be unconditional security recently.[37]

When there is noise in the channel, an adversary Eve can gain a certain amount of information by hiding her presence in the channel noise. In these cases, the information leakage should be eliminated by using either quantum error correction[38] or quantum privacy amplification.[39] However, quantum privacy amplification is complex to implement, and it also ruins the direct communication picture as it involves the merger and order reshuffling of photons. An efficient way to implement QSDC in noisy channel is to use quantum error correction.[38,40] As a matter of fact, post-processing can be performed by using quantum error correction without using privacy amplification and reconciliation in QKD.[41] In this work, instead of using a complicated quantum error correction scheme, we introduce a simple frequency coding scheme into the DL04 protocol and propose new QSDC protocol, which we call simple-coded DL04 (SICO-DL04) protocol. In SICO-DL04 protocol, the information is encoded on the spectrum of a sequence of single photons rather than on the single qubits in the original DL04 protocol. SICO-DL04 can work efficiently in the presence of channel loss and noise. We have demonstrated the SICO-DL04 protocol experimentally.

**MATERIALS AND METHODS**

Here we describe the SICO-DL04 protocol briefly. Suppose that Bob wants to send secret information to Alice. The protocol contains the following four steps:

1) Alice prepares a sequence of $N_2$ single photons. Each photon of the sequence is randomly in one of the four states $|0\rangle$, $|1\rangle$, $|+\rangle$, and $|-\rangle$, where $|0\rangle$ and $|1\rangle$ are the eigenstates of the Pauli **Z** operator, and $|\pm\rangle = (|0\rangle \pm |1\rangle)/\sqrt{2}$ are the eigenstates of the Pauli **X** operator. Then Alice sends the photon sequence to Bob, Bob acknowledges this fact.

2) Because of channel noise and loss, Bob receives only $N_1$ single photons ($N_1<N_2$). He selects $CN_1$ number ($C$ is a positive number less or equal to 1/2) of photons randomly from the $N_1$ received photons for eavesdropping check by measuring them randomly in the X-basis or the Z-basis. Bob tells Alice the positions, the measuring-basis and the measuring results of these measured photons. Alice compares her results with those of Bob and obtains an error rate. If the error rate is higher than the threshold, they abort the communication. If the error rate is less than the threshold, the Alice-Bob communication is considered as safe, and the communication continues to step 3.

3) The remaining $(1-C)N_1$ received photons are used for encoding the secret information. He also selects $C(1-C)N_1$ single photons as check bits for the Bob-Alice transmission, and applies randomly one of the two operations, $I = |0\rangle\langle0|+|1\rangle\langle1|$, and $U = i\sigma_y=|0\rangle\langle1|-|1\rangle\langle0|$, which does nothing or flip the state of the photon. The rest of the single photons will be processed by a simple frequency coding scheme, which will be described later.

4) Bob sends the encoded photon sequence back to Alice who can deterministically decode the

qubits by measuring the photons in the same basis she prepared, without demand for a classical channel. Alice gets the bit value of each single photon in the sequence and their arrival time. Because of channel loss, Bob receives only $N$ ($N \leq (1-C)^2 N_1$) photons in each block after subtracting the check photons. Alice and Bob will also publicly compare the results of the checking bits to ensure if there exists eavesdropping in the Bob-Alice transmission. Next, Alice calculates the spectrum and determines Bob's encoded bits and reads out the secret information.

Now we describe the **simple frequency coding** scheme. In the DL04 protocol, the information is directly encoded on the single photon state, where 0 is encoded with $I$ and 1 with $U = i\sigma_y = |0\rangle\langle 1| - |1\rangle\langle 0|$ respectively. The operation $U$ flips the state without changing the measurement basis, namely,

$$U|0\rangle = -|1\rangle, \quad U|1\rangle = |0\rangle,$$

$$U|+\rangle = |-\rangle, \quad U|-\rangle = -|+\rangle. \tag{1}$$

However, in a noisy channel, such a direct coding is difficult due to noise and loss. Instead of using a single operation to encode a bit value, simple frequency coding uses a series of periodic operations on a photon sequence to encode information. Bob flips the state in a sequence of photons according to a periodic function with period $T=1/f$, where $f$ is the modulation frequency that encodes the information. Once Alice obtains the modulation frequency after she measures a sequence of single photons, she gets Bob's information fully. The modulation signal Bob applies to flip the photon sequence, after excluding the checking bits, is

$$\text{Operation} = \begin{cases} \text{flip}, & \delta_T + 2nT < \tau_i \leq (2n+1)T + \delta_T, \\ \text{no flip}, & \delta_T + (2n+1)T < \tau_i \leq (2n+2)T + \delta_T, \end{cases} \tag{2}$$

where $T$ is the modulation signal period. Different $T$ encodes different bit values. $\delta_T$ is an phase offset for each modulating signal, and it is random so that Eve could not guess what the period $T$. The measured value that Alice obtains is denoted as 0 if there is no state change, and 1 if there is state flip. She also records the arrival time $\tau_i$, for $i=1, 2, 3,…, N$, where $N$ is the number of optical photons that Alice measured in each block after subtracting the check photons. An illustrating example is given in Table 1, where the initial states, the final states, the measured value $x_{(i)}$ and arrival time $\tau_i$ are shown.

Table 1. Polarization modulation of single photons

| Initial state | $\updownarrow$ | $\nearrow$ | $\updownarrow$ | $\searrow$ | $\leftrightarrow$ | ... | $\leftrightarrow$ |
|---|---|---|---|---|---|---|---|
| Final state | $\updownarrow$ | $\searrow$ | $\leftrightarrow$ | $\searrow$ | $\updownarrow$ | ... | $\leftrightarrow$ |
| $x_{(i)}$ | 0 | 1 | 1 | 0 | 1 | ... | 0 |
| Time $\tau_i$ | $\tau_1$ | $\tau_2$ | $\tau_3$ | $\tau_4$ | $\tau_5$ | ... | $\tau_N$ |

Of course, not all the photons can arrive and be detected by Alice because of the loss of optical fibre and imperfect detection efficiency of single photon detector. Then nothing will be recorded for the lost photons. This simple frequency coding scheme is robust against error and loss. The coding frequency can be accurately determined from the sequence ($x_{(i)}$, $\tau_i$) using the discrete time Fourier transform,

$$X_{(f)} = \sum_{i=1}^{N} x_{(i)} e^{-j \cdot 2\pi f \tau_i} . \quad (3)$$

This spectrum will have a peak at the coding frequency $f=1/T$. From the spectrum, Alice can easily determine the coding frequency, hence reads out the encoded information.

For a given quantum communication system, there exists a finite maximum number $N_c$ of frequency channels,

$$N_c = \frac{f_{max} - f_{min}}{f_b} + 1, \quad (4)$$

where $f_{max}$ and $f_{min}$ are the maximum and minimum modulation frequencies respectively; $f_b$ is the channel frequency spacing. If Bob uses only a single frequency to modulate one single photon sequence, then each block can send $N_c$ values. However, the capacity can be enlarged considerably if Bob loads more than one frequency component onto the photon sequence simultaneously. Assuming Bob loads $r$ frequency components on one single photon sequence. The effective degrees of freedom are the total number of different combinations of $r$ frequencies,

$$N_{max} = \frac{N_C!}{r!(N_C - r)!}, \tag{5}$$

which means one single photon sequence can carry $b=\log_2 N_{max}$ bits of information. The transmission rate can be expressed as

$$I = \frac{b}{T_{span}} = \frac{1}{T_{span}}\log_2 N_{max}, \tag{6}$$

where $T_{span}$ is the time span, i.e., the time length of photon sequence. In an example with a 500 MHz modulation bandwidth, $f_b$=1.5 kHz, $T_{span}$ =1$ms$ and frequency components $r$=3, the transmission rate can reach $I$=42.5 kbps. This coding scheme is similar to the ultra-wide-band communication in the field of wireless communication.[42] With more sophisticated coding algorithms, the system performance could be further improved.

**RESULTS AND DISCUSSION**

The experimental setup is shown in Fig.1. The encoding process is controlled by a field programmable gate array (FPGA) device. In order to guarantee the security of the secret information, the eavesdropping detection process must be finished before the encoding procedure. During the eavesdropping detection procedure of the block, an optical fibre (with length $L_2$) is used as a delay line to store the encoded photons. In practical application, single photons are approximated by attenuated weak light pulses. The photon number of coherent light pulse obeys Poisson distribution. Because of this, there is a white noise in the spectrum of the measured sequence. The bandwidth of the modulation frequency used in our experiment is about 400 kHz.[43] In our experiment, we take

$N_c$=16 frequency channels within the 400 kHz bandwidth and frequency spacing of 25 kHz. The results of the spectral analysis of the sequence ($x_{(i)}$, $\tau_i$) using Eq. (3) are shown in Fig. 2. It shows that there is a peak at each of the modulation frequency among the white noise background. All together there are 16 such peaks. Though the noise and loss of the quantum channel broaden the peak and reduce the height of peak, the central frequency remains the same, which determines the coding frequency accurately. This enables Alice to read out the information encoded by Bob. The identification of the modulation frequency is possible only when the signal-to-noise is higher than 1. Figure 3 shows the relationship between the signal and background noise with different mean photon numbers. It indicates that with a bigger relative photon number per pulse, the modulation signal is much higher than the background noise. We also draw the background noise distribution in Figure 3. It is shown that as the mean photon number per pulse increases, the background noise remains low. In our frequency coding experiment, the system repetition frequency is 10 MHz. The number of photons used for spectrum calculation in a block is $N$=80. The time span of the single photon block is $T_{span}$ =1$ms$. We take $r$=1, hence only one frequency channel will respond to the received signal at any given moment, which means Alice can get $\log_2 16$=4 bits of information by processing one block of data in one time span. The communication rate in the experiment is 4 kbps.

**Now we discuss the security of the SICO-DL04 protocol briefly**. Here we give a brief analysis of the security of SICO-DL04. These results will give a rough estimate and rigorous unconditional security proof of the protocol will not be presented in this paper. As we have discussed before, no one can acquire the frequency information unless the length of the sequence ($x_{(i)}$, $\tau_i$) they collected is long enough for spectrum calculation. If the mean number of qubits per pulse Eve obtained, $R_{Eve}$, is small enough, Eve cannot distinguish the modulation frequency from background noise. So, she would not get any useful information. For a given quantum communication system, once all the parameters such as mean photon number, communication distance, etc. have been set, the amount of information that the receiver (or an eavesdropper) can obtain is predictable. When the single-photon source is an attenuated laser pulse with mean photon number $\mu$ per pulse, the probability to have $n$ photons in a single pulse follows the Poisson distribution $p_n$. The probability that an optical pulse could be detected at the receiving end is[44]

$$P = \sum_{n\geq 1} p_n [1-(1-\eta_{det}\eta(1-C))^n] \approx \eta_{det}\eta(1-C)\mu, \qquad (7)$$

where $\eta=10^{-\alpha(2L_1+L_2)/10}$ is the optical attenuation due to the loss of the fiber (the total fiber length is $2L_1+L_2$); $L_1$ and $L_2$ are the communication distance and the optical delay line at Bob side respectively. $\alpha$ is the optical fiber loss coefficient (typical value is 0.2 dB/km); $\eta_{det}$ is the quantum efficiency of the single photon detector.[45,46] Equation (7) is valid if $\eta_{det}\,\eta\,p_n\,n \ll 1$ for all $n$. With multiphoton pulses, Eve performs a quantum nondemolition measurement on the pulses as soon as they exit Alice's station. When $n=2$ Eve stores one photon and sends the other one to Bob by using a lossless channel. After Bob's encoding operation, Eve captures the photon again. In order to gain Bob's secret information, Eve must judge whether the polarizations of the two photons are parallel or antiparallel. As described in Ref. 47, there exists an optimal measurement which gives Eve a conclusive result with probability 1/4. Eve discards all the inconclusive results and retains the conclusive ones. The mean amount of effective photons that Eve gets within an operation-cycle is

$$R_{Eve}^{n=2} = \frac{1}{4}[10^{-\alpha L_2/10} p_2(1-C)], \qquad (8)$$

where $p_2$ is the Poisson distribution component for $n=2$. When $n=3$, there exists a measurement $M$ that provides a conclusive result about whether the polarization is flipped or not with a probability 1/2 (ref. 48). For pulses with three or more photons, she executes $M$, if the outcome is not conclusive she blocks these pulses, and if the outcome is conclusive she prepares a new photon in the same state and forwards it to Bob. After Bob's encoding operation, Eve measures the photon again on the backward trip to see whether the polarization state has been flipped. From these operations, the mean amount of effective qubits that Eve can get is

$$R_{\text{Eve}}^{n\geq 3} = \frac{1}{2}[10^{-\alpha L_2/10}(1-C)\sum_{n=3}^{\infty} p_n] . \qquad (9)$$

Both eavesdropping methods do not cause any bit error, which means Eve could not be detected during such an eavesdropping process.

In a noisy channel, Eve may also gain a certain amount of data without being detected by hiding her presence in the noise if she replaces the noisy channel by an ideal one and sends another photon prepared by herself to Alice. She could acquire a fraction $4e$ of the qubits on the forward Alice-Bob channel, where $e$ is the bit error rate caused by channel noise. The factor 4 arises because there is a 50% chance for Eve to pick the correct basis, and for those she picks the wrong basis she has another 50% chance of not causing a bit error. The mean qubits that Eve can get is

$$R_{\text{Eve}}^{n=1} = 4[10^{-\alpha(L_1+L_2)/10} p_1 e(1-C)] . \qquad (10)$$

Considering all the strategies, the mean number of photons that Eve eventually gets is

$$R_{\text{Eve}} = R_{\text{Eve}}^{n=1} + R_{\text{Eve}}^{n=2} + R_{\text{Eve}}^{n\geq 3} . \qquad (11)$$

The number of photons that Alice gets and the *transmission rate* of Alice respectively could be derived from equation (6)

$$R_{\text{Alice}} = 10^{-\alpha(2L_1+L_2)/10} \eta_{\text{det}} \mu (1-C) , \qquad (12)$$

$$I_{\text{Alice}} = \frac{bR_{\text{Alice}}}{NT_{\text{span}}} , \qquad (13)$$

where $b$ is the bit-string length of the secret information encoded in a single photon sequence, and $N$ is the number of qubits that Alice received from a single photon sequence for information recovery in one time span. The identification of the modulation frequency needs enough data for spectrum calculation. Although Eve gets some qubits, it does not mean that she can get any *information bit*. From the viewpoint of information theory,[49] Eve cannot get the *information bits* more than $b$ when the number of qubits she gets is smaller than $b$. Therefore, considering the equations (12~13), the condition of security is $R_{\text{Alice}} / R_{\text{Eve}} > N / b$. The secure *information bits* per pulse and secure communication distance were shown in Figure 4. The distance depends on the intensity of laser pulses. For weak laser pulses, the secure distance is about 10 kilometers, which is similar to two-way QKD with weak laser pulses.

**CONCLUSIONS**

In summary, we presented a new practical QSDC protocol based on single photons, the SICO-DL04 protocol, in which a novel quantum frequency coding scheme was developed. In the protocol, instead of encoding the bit values on the individual photons, the information is encoded in the modulation frequencies which flip the states of photons in a sequence periodically. Because the information is encoded on the statistical properties of the sequence, it is robust against channel loss and noise. The SICO-DL04 protocol does not require privacy amplification procedure which ruins the quantum secure direct communication picture. The frequency coding scheme can also adopt several modulation frequencies simultaneously, increasing considerably the amount of information that a block of single photons can carry. It is also simpler than protocols encoded with complicated quantum error correction code. The security of the protocol is also carefully analyzed, taking into major eavesdropping attack strategies and channel noise and loss.

We have demonstrated the SICO-DL04 protocol experimentally using existing technology. This is the first time that the block transmission has been demonstrated, hence the first experimental demonstration of QSDC. In our experiment, the transmission is carried out in blocks, each block contains 80 photons. The range of the modulation frequency is 400 kHz, and frequency spacing is 25 kHz. Each single photon sequence can transmit one of the 16 frequency values, which is 4 bits of information. A transmission rate of 4 kbps has been demonstrated in the experiment. The experiment has firmly demonstrated the principle of QSDC with current technology and practical channel noise and loss.

**ACKNOWLEDGMENTS**

The project is sponsored by 973 Program (2012CB921603), 863 Program (2011AA010801), the Natural Science Foundation of China (11174187, 10934004 and 11204166), the Doctoral Foundation of the Education Ministry of China (20121401120016) and PCSIRT (IRT 13076). Qin GQ and Long GL are supported by National Natural Science Foundation of China under Grant Nos. 11175094 and 91221205, and the National Basic Research Program of China under Grant No. 2015CB921001.

______________________________________________________

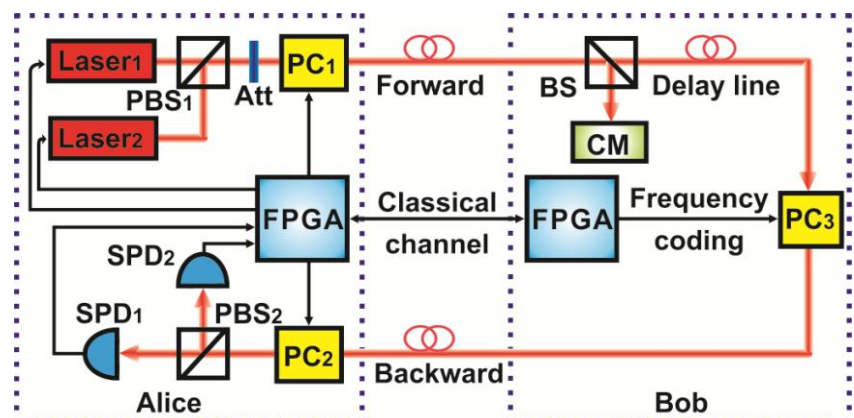

Figure 1 Schematic diagram of experimental setup of SICO-DL04 protocol. PBS: Polarization beam splitter; Att: Variable attenuator; PC: Polarization controller; BS: Beam splitter; CM: Control model; FPGA: Field Programmable Gate Array; SPD: Single photon detector. The distance between Alice and Bob is $L_1$, and the delay line length is $L_2$.

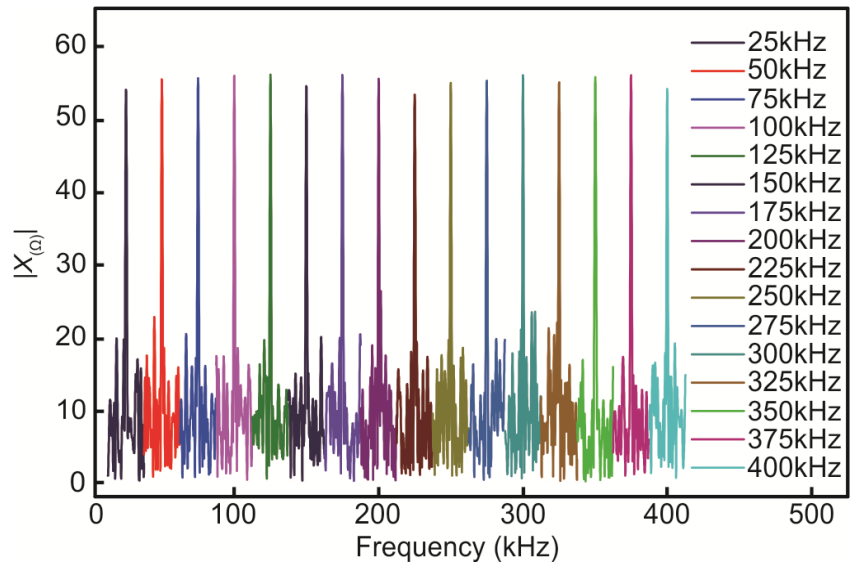

Figure 2 The modulation spectrum. The *y*-axis is the Fourier transformed amplitude in Eq.(3). The different colors represent different modulation frequencies, hence different values of information. If each block of photons is modulated by a single frequency, then it carries $2^4$=16 values, or 4 bits. The background white noise is due to the imperfect single photon detector, channel noise.

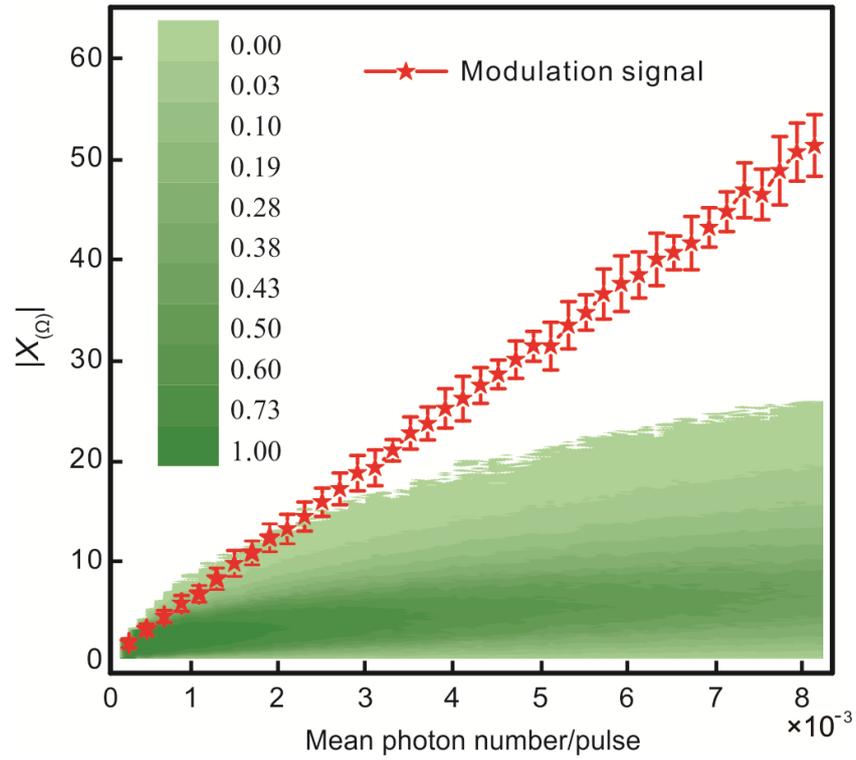

Figure 3 The signal and background noise distribution of the modulation spectrum. The *x*-axis is the average number of photon counts per pulse that Alice detects. The colored areas are the distribution of the background noises, where the colors represent the relative probability of the noises with the respective amplitudes. The red line represents the amplitude of the modulation peak. The modulation frequency is chosen as 200 kHz. The system repetition frequency is 10 MHz.

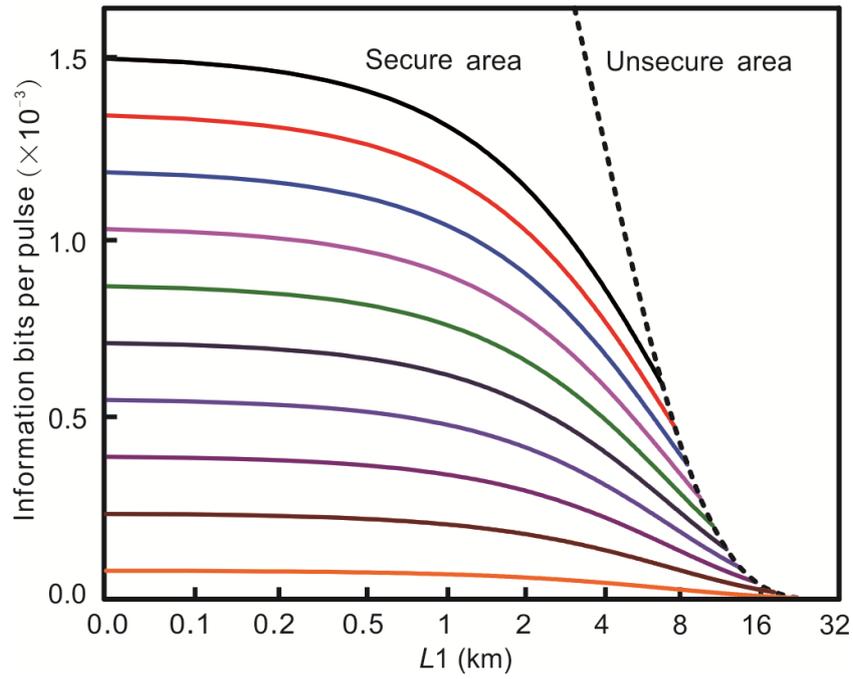

Figure 4 The transmission bit per pulse versus the communication distance. The dotted line is the cut-off line of the secure communication area. The solid lines with different color represent different mean photon number per pulse ($\mu$=0.19, 0.17, 0.15, 0.13, 0.11, 0.9, 0.7, 0.5, 0.3, 0.1, from top to bottom). Here $\eta_{det}$=0.32, $e$=0.5%, $\alpha$=0.2 dB/km, $L_2$=$L_1$, $C$=1/2.